\documentclass[journal]{IEEEtran}
\ifCLASSINFOpdf
\else
\fi
\usepackage{stfloats}
\usepackage{amsmath,amsfonts,amssymb,amscd,amsthm}
\usepackage{graphicx,epsfig}
\usepackage{subcaption}
\usepackage{cases}
\usepackage[font=singlespacing]{caption}
\usepackage{lscape}
\usepackage{verbatim}
\usepackage{epstopdf}
\usepackage{array}
\usepackage{balance}
\usepackage{multirow}
\usepackage[noadjust]{cite}
\usepackage{bbold}
\usepackage{multirow}
\usepackage{url}

\theoremstyle{definition}
\newtheorem{example}{Example}
\newtheorem{remark}{Remark}
\newcommand{\qedblack}{\hfill \ensuremath{\blacksquare}}

\begin{document}
\title{Power Divider}
\author{Yu Christine Chen,~\IEEEmembership{Member,~IEEE} and Sairaj V.~Dhople,~\IEEEmembership{Member,~IEEE}
\thanks{Y.~C.~Chen is with the Department of Electrical and Computer Engineering at the University of British Columbia, Vancouver, Canada.  E-mail: \texttt{chen@ECE.UBC.EDU}.}
\thanks{S.~V.~Dhople is with the Department of Electrical and Computer Engineering at the University of Minnesota, Minneapolis, MN, USA.  E-mail: \texttt{sdhople@UMN.EDU}.}%His work is supported in part by the Minnesota's Discovery, Research and Innovation Economy grant we well as by the National Science Foundation through the CAREER award 1453921.
}
\maketitle
\begin{abstract}
\boldmath
This paper derives analytical closed-form expressions that uncover the contributions of nodal active- and reactive-power injections to the active- and reactive-power flows on transmission lines in an AC electrical network. Paying due homage to current- and voltage-divider laws that are similar in spirit, we baptize these as the~\emph{power divider} laws. Derived from a circuit-theoretic examination of AC power-flow expressions, the constitution of the power divider laws reflects the topology and voltage profile of the network. We demonstrate the utility of the power divider laws to the analysis of power networks by highlighting applications to transmission-network allocation, transmission-loss allocation, and identifying feasible injections while respecting line active-power flow set points.
\end{abstract}

\IEEEpeerreviewmaketitle

\section{Introduction} 
\label{sec:intro}

\begin{figure}[t!]
\centering
\mbox{
\epsfig{file=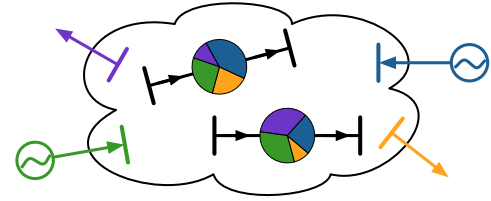,width=0.36\textwidth}}
\caption{The power divider laws: mapping injections to flows.}
\label{fig:cloud}
\vspace{-14pt}
\end{figure}

\IEEEPARstart{T}{his} paper presents analytical closed-form expressions that map nodal active- and reactive-power injections to active- and reactive-power flows on transmission lines in a power network operating in sinusoidal steady state (see Fig.~\ref{fig:cloud} for an illustration). We term these the \emph{power divider laws}, since their form and function are analogous to and reminiscent of voltage- and current-divider laws that are widely used in circuit analysis~\cite{Sedra-2007}.\footnote{While it should be contextually clear, it is worth pointing out that passive devices that route power in microwave engineering applications are also called \emph{power dividers}~\cite{Wilkinson-1960}.}

Arguably, the most obvious application of this work would be in transmission-network allocation, where one seeks to equitably and systematically apportion the cost of loading transmission lines in a power system to constituent generators and loads.  Indeed, we demonstrate how the power divider laws can be leveraged for this task. More generally, quantification of the impact of injections on flows is germane to numerous applications in power-system analysis, operation, and control. (We interchangeably refer to nodal complex-power injections and line complex-power flows simply as \emph{injections} and \emph{flows}, respectively. Furthermore, we  implicitly consider \emph{injections} to be positive for generation and negative for load.) In this paper, we provide a snapshot of additional possibilities by focusing on the particular tasks of allocating transmission-line losses and identifying feasible injections under the constraint of line active-power flow set points.  Through this promising (but by no means exhaustive) set of examples, we demonstrate how the power divider laws yield analytical insights, provide computational benefits, and improve accuracy compared to state-of-the-art approaches.  For instance, with regard to the task on identifying feasible injections, using the power divider laws we formulate a linearly constrained least-squares problem, the solution of which returns the set of injections that respect line active-power flow set points. On the other hand, with regard to loss allocation, we recover exact expressions that map transmission-line losses to  bus active- and reactive-power injections.  This has been recognized to be a challenging problem in the literature~\cite{Conejo:2002, Ding:2006}.  Consider, e.g., the following remark from~\cite{Conejo-2001}:\emph{``[...] system transmission losses are a nonseparable, nonlinear function of the bus power injections  which makes it impossible to divide the system losses into the sum of terms, each one uniquely attributable to a generation or load.''} In this paper, using the power divider laws, we demonstrate how losses on an \emph{individual transmission line} can be~\emph{uniquely} attributed to each generator and load.

Given the obvious utility of mapping injections to flows in a variety of applications pertaining to power system operation and control, many algorithmic approaches and approximations have been applied to tackle this problem in the literature. Some relevant prior art in this regard include: numerical approaches~\cite{Kirschen-1997, Bialek:1998}, integration-based methods~\cite{Fradi-2001}, utilizing current flows as proxies for power flows~\cite{Conejo-2007}, and leveraging generation shift distribution factors~\cite{Rudnick:1995, daSilva:2013}. Distinct from the numerical methods in~\cite{Kirschen-1997,Fradi-2001,Bialek:1998, Conejo-2007}, the power divider laws offer an \emph{analytical} approach to precisely quantify the relationship between line flows and nodal injections. Furthermore, distinct from~\cite{Rudnick:1995, daSilva:2013}, our approach tackles the nonlinear power-flow expressions and acknowledges injections from \emph{all} buses in the network; no assumption is made about the existence or location of a slack bus that makes up for power imbalances.

The derivation of the power divider laws begins with an examination of the algebraic power-balance expressions in matrix-vector form. From these, we extract the exact nonlinear mapping between line complex-power flows and nodal complex-power injections in analytical closed form. The first step proceeds in the same vein as~\cite{Conejo-2007}, and it involves uncovering the contributions of nodal current injections to line-current flows. It emerges that this is a linear function  of entries of the network admittance matrix and independent of the network's voltage profile, i.e., bus-voltage magnitudes and phase angle differences. We refer to the sensitivities of line current flows to nodal current injections as \emph{current injection sensitivity factors}. Leveraging this first step, we then derive the impact of bus complex-power injections on line complex-power flows through algebraic manipulations of the power-flow expressions written in matrix-vector form. The resultant nonlinear expressions that constitute the power divider laws are  functions of the voltage profile of the AC electrical network, as well as network topology (since they include the current injection sensitivity factors by construction). Additionally, we employ engineering insights on small angle differences, a uniform voltage magnitude profile, and the inductive nature of transmission networks to obtain simplified approximations of the power divider laws. For instance, one such simplification decouples the active- and reactive-power flows and injections. Under the most restrictive set of assumptions, power divider laws yield the ubiquitous DC power flow expressions~\cite{Glover:2012}.

The remainder of this manuscript is organized as follows.  Section~\ref{sec:prelim} establishes mathematical notation and the power system model. In Section~\ref{sec:line}, beginning with current injection sensitivity factors, we outline the derivation of the power divider laws. Approximations that emphasize decoupling of active- and reactive-power flows and injections as well as connections to the canonical DC power flow are provided in Section~\ref{sec:approx}.  Section~\ref{sec:app} highlights applications of the power divider laws to transmission-network allocation, loss allocation, and feasible-injection identification.  These applications are illustrated via numerical case studies in Section~\ref{sec:cases}. Finally, concluding remarks and directions for future work are provided in Section~\ref{sec:conc}. 

\section{Preliminaries} \label{sec:prelim}
In this section, we  introduce relevant notation and describe the power system model used in the remainder of the paper.  
\subsection{Notation} \label{sec:Notation}
The matrix transpose is denoted by $(\cdot)^\mathrm{T}$, magnitude of a complex number by $|\cdot|$; complex conjugate by $(\cdot)^*$, complex-conjugate transposition by $(\cdot)^\mathrm{H}$, real and imaginary parts of a complex number or vector by $\mathrm{Re}\{\cdot\}$ and $\mathrm{Im}\{\cdot\}$, respectively, and $\mathrm{j} := \sqrt{-1}$. A diagonal matrix formed with entries of the vector $x$ is denoted by $\mathrm{diag}(x)$; and $\mathrm{diag}(x/y)$ forms a diagonal matrix with the $m$-th entry given by $x_m/y_m$, where $x_m$ and $y_m$ are the $m$-th entries of vectors $x$ and $y$, respectively. The spaces of $N\times1$ real- and complex-valued vectors are denoted by $\mathbb{R}^N$ and $\mathbb{C}^N$, respectively, and $\mathbb{T}^{N}$ denotes the $N$-dimensional torus.  The spaces of $M\times N$ real- and complex-valued matrices are denoted by $\mathbb{R}^{M\times N}$ and $\mathbb{C}^{M\times N}$, respectively.  The entry in the $m$-th row and $n$-th column of the matrix $X$ is denoted by $[X]_{mn}$. The $N \times 1$ vectors with all ones and all zeros are denoted by $\mathbb{1}_{N}$ and $\mathbb{0}_{N}$; $e_m$ denotes a column vector of all zeros except with the $m$-th entry equal to~$1$; and $e_{mn} := e_m - e_n$. For a vector $x=[x_1,\dots,x_N]^\mathrm T$, we denote $\cos(x):= [\cos(x_1),\dots,\cos(x_N)]^\mathrm T$, $\sin(x):= [\sin(x_1),\dots,\sin(x_N)]^\mathrm T$, and $\mathrm{exp}(x):= [\mathrm e^{x_1},\dots,\mathrm e^{x_N}]^\mathrm T$. Finally, $\Pi_m := [e_1, \dots, e_{m-1}, e_{m+1}, \dots, e_{N}] \in \mathbb{R}^{N \times N-1}$. Corresponding to the vector $x \in \mathbb{C}^N$, we denote $\widetilde x^m  := \Pi_{m}^\mathrm T x$; note that $\widetilde x^m$ is recovered from the vector $x$ by removing the $m$-th entry.

\subsection{Power System Network Model}
Consider a power system with $N$ buses collected in the set $\mathcal{N}$.  Lines in the electrical network are represented by the set of edges $\mathcal{E} := \{(m,n)\} \subset \mathcal{N} \times \mathcal{N}$. Denote the bus admittance matrix by $Y = G + \mathrm{j} B \in \mathbb{C}^{N\times N}$. Transmission line $(m,n)\in \mathcal{E}$ is modelled using the lumped-element $\Pi$-model with series admittance $y_{mn} = y_{nm} = g_{mn} + \mathrm{j} b_{mn}\in \mathbb{C} \setminus\{0\}$ and shunt admittance $y_{mn}^{\mathrm{sh}} = g_{mn}^{\mathrm{sh}} + \mathrm j b_{mn}^{\mathrm{sh}} \in \mathbb{C}\setminus\{0\}$ on both ends of the line. Entries of the admittance matrix are defined as
\begin{equation}
[Y]_{mn} := \left \{\begin{array}{ll} 
{y}_{m} + \sum_{(m,k) \in \mathcal{E}} y_{mk} , & \textrm{if } m=n,\\ 
- y_{mn} , &\textrm{if }(m,n)\in \mathcal{E},\\
0, & \textrm{otherwise}, \end{array} \right.
\label{eq:Ymatrix}
\end{equation}
where 
\begin{equation}
y_{m} = g_m + \mathrm{j}b_m := y_{mm} + \sum_{k \in \mathcal{N}_m} y_{mk}^{\mathrm{sh}},
\end{equation}
denotes the total shunt admittance connected to bus $m$ with $\mathcal{N}_m \subseteq \mathcal{N}$ representing the set of neighbours of bus $m$ and $y_{mm}\in \mathbb{C}$ any passive shunt elements connected to bus $m$.  Notice that if the electrical network had no shunt elements, then $Y$ is a singular matrix with zero row and column sums. On the other hand, the inclusion of (even one) shunt admittance term in~\eqref{eq:Ymatrix} intrinsically guarantees invertibility of $Y$ by rendering it irreducibly diagonally dominant~\cite{Horn:2013}.

Let $V = [V_1, \dots, V_N]^\mathrm{T}$, where $V_i = |V_i| \angle \theta_i \in \mathbb{C}$ represents the voltage phasor at bus $i$, and let $I = [I_1, \dots, I_N]^\mathrm{T}$, where $I_i \in \mathbb{C}$ denotes the phasor of the current injected into bus $i$. The bus-voltage magnitudes are collected in the vector $|V| = [|V_1|,\dots,|V_n|]^\mathrm T \in \mathbb{R}_{>0}^N$, and bus-voltage angles are collected in the vector $\theta = [\theta_1,\dots,\theta_N]^\mathrm T \in \mathbb{T}^N$. Kirchhoff's current law for the buses in the power system can be compactly represented in matrix-vector form as
\begin{equation}
I =  Y V.
\label{eq:Ii}
\end{equation}
Denote the vector of bus complex-power injections by $S = [S_1, \dots, S_N]^\mathrm T = P + \mathrm{j} Q$, with $P = [P_1, \dots, P_N]^\mathrm T$ and $Q = [Q_1, \dots, Q_N]^\mathrm T$. (By convention, $P_i$ and $Q_i$ are positive for generation and negative for loads.)  Then, bus complex-power injections can be compactly written as
\begin{equation}
S  =  \mathrm{diag}\left(V\right) I^*.
\label{eq:S}
\end{equation}
The equation above is the complex-valued equivalent of the standard power flow equations expressed in a compact matrix-vector form, and generalized to include active- and reactive-power injections as well as voltage magnitudes and phase angles at all buses (including the artifactual slack bus).     
We will find the following auxiliary bus voltage phase angle- and active power-related variables useful:
\begin{equation} \label{eq:theta_defs}
\theta^m := \theta_{m} \mathbb{1}_{N} - \theta, \quad \widetilde \theta^m := \Pi_{m}^\mathrm T \theta^m, \quad \widetilde P^m:= \Pi_{m}^\mathrm T P,
\end{equation}
where $\theta^m$ is obtained by setting the system angle reference as $\theta_{m}$ (the voltage phase angle at bus $m$).  In~\eqref{eq:theta_defs}, $\widetilde \theta^m$ and $\widetilde P^m$ denote the $(N-1)$-dimensional vectors that result from removing the $m$-th entries from the $N$-dimensional vectors $\theta^m$ and $P$ respectively.  For example, with respect to the system illustrated in~Fig.~\ref{fig:3bus}, where $N=3$, relevant variables are $\theta = [\theta_1, \theta_2, \theta_3]^\mathrm T$, $P = [P_1, P_2, P_3]^\mathrm T$.  With the choice  $m=2$, it follows that $\theta^{2} = [\theta_2 - \theta_1, 0, \theta_2 - \theta_3]^\mathrm T$, $\widetilde \theta^2 = [\theta_2 - \theta_1, \theta_2 - \theta_3]^\mathrm T$, and $\widetilde P^2 = [P_1,P_3]^\mathrm T$.

\section{Derivation of the Power Divider Laws}
\label{sec:line}

In this section, we demonstrate how line current flows can be expressed as a linear function of bus current injection phasors based on entries of the network admittance matrix.  Leveraging this relationship, we then derive expressions that uncover the contributions of bus active- and reactive-power  injections to line active- and reactive-power flows.

\subsection{Line Current Flows}
We begin by expressing the current in line $(m,n) \in \mathcal{E}$ as
\begin{align}
I_{(m,n)} &= y_{mn} (V_m - V_n)  + y_m V_m \nonumber \\
& = \left( y_{mn} e_{mn}^\mathrm{T} + y_m e_{m}^\mathrm{T}   \right)  V.
\label{eq:Ikl_mtx}
\end{align}
Since the bus admittance matrix, $Y$, is invertible, from~\eqref{eq:Ii}, the bus voltages can be expressed as $V =  Y^{-1}  I$. As a direct consequence,~\eqref{eq:Ikl_mtx} can be written as 
\begin{equation}
I_{(m,n)} = \left(  y_{mn} e_{mn}^\mathrm{T}  + y_m e_{m}^\mathrm{T}   \right) Y^{-1} I =: \kappa_{(m,n)}^\mathrm{T}  I,
\label{eq:amn_inv}
\end{equation}
where $\kappa_{(m,n)} \in \mathbb{C}^{N}$. The entries of $\kappa_{(m,n)}$ are referred to as the~\emph{current injection sensitivity factors} of line $(m,n)$ with respect to the bus current injections. The current injection sensitivity factors in~\eqref{eq:amn_inv} delineate the exact effect of bus current injections on the current in line~$(m,n)$.  Moreover, they depend only on network parameters, and not the operating point, i.e., the voltage magnitudes and phase angles across the network do not influence~\eqref{eq:amn_inv}. For subsequent developments, we will find it useful to decompose the current injection sensitivity factors into real and imaginary components as 
\begin{equation} \label{eq:a}
\kappa_{(m,n)} = \alpha_{(m,n)} + \mathrm{j} \beta_{(m,n)}.
\end{equation} 
Note that the expression in~\eqref{eq:amn_inv} is a generalization of the~\emph{current divider} law, which typically applies to the particular case of a single current source feeding a set of impedances connected in parallel~\cite{Sedra-2007}.

\begin{remark}[Invertibility of $Y$]
\label{rem:Yinv}
In~\eqref{eq:amn_inv}, invertibility of $Y$ is implicitly assumed.   On the other hand, suppose $Y$ is not invertible, i.e., there are no ground-connecting shunt elements. The current in line $(m,n)$ is then given by
\begin{equation}
I_{(m,n)} = y_{mn} (V_m - V_n) = y_{mn} e_{mn}^\mathrm{T} V.
\label{eq:Ikl}
\end{equation}
In this case, the bus voltages satisfy
\begin{equation} \label{eq:V}
V = Y^\dagger I + \frac{1}{N}\mathbb{1}_{N}\mathbb{1}_{N}^\mathrm{T}V,
\end{equation} 
which follows from pre-multiplying both sides of \eqref{eq:Ii} by the pseudoinverse of the admittance matrix, $Y^\dagger$, and recognizing that the admittance matrix and its pseudoinverse are related by~\cite{Horn:2013}
\begin{equation*} \label{eq:YpinvY}
Y Y^\dagger = Y^\dagger Y = \mathrm{diag}(\mathbb{1}_N) - \frac{1}{N}\mathbb{1}_{N}\mathbb{1}_{N}^\mathrm{T}.
\end{equation*}
Substituting for $V$ from~\eqref{eq:V} in~\eqref{eq:Ikl}, we see that the current injection sensitivity factors in this case are given by
\begin{equation}
\kappa_{(m,n)}^\mathrm{T} =   y_{mn} e_{mn}^\mathrm{T} Y^{\dagger}
\label{eq:amn_noninv}
\end{equation}
which follows from the fact that $e_{mn}^\mathrm{T} \mathbb{1}_{N}\mathbb{1}_{N}^\mathrm{T}= \mathbb{0}_{N}^\mathrm T$. \qedblack
\end{remark}

\subsection{Line Active- and Reactive-power Flows}
\label{sec:powerdiv}
Denote, by $S_{(m,n)} = P_{(m,n)} + \mathrm{j}Q_{(m,n)}$, the complex power flowing across line $(m,n)$.  We can write
\begin{equation}
S_{(m,n)}  =  V_m I_{(m,n)}^*.
\label{eq:Smn1}
\end{equation}
We substitute the current injection sensitivity factors from~\eqref{eq:amn_inv} (or~\eqref{eq:amn_noninv} for non-invertible admittance matrices) into~\eqref{eq:Smn1}, and obtain
\begin{equation}
S_{(m,n)}  =  V_m \kappa_{(m,n)}^\mathrm{H} I^*.
\label{eq:Smn}
\end{equation}
Eliminating $I^*$ from~\eqref{eq:Smn} using~\eqref{eq:S}, we get 
\begin{equation}
S_{(m,n)} = V_m \kappa_{(m,n)}^\mathrm{H} \left(\mathrm{diag}\left(V\right)\right)^{-1} S.
\label{eq:Smn3}
\end{equation}
With the phasor form of the voltages, we can write~\eqref{eq:Smn3} as
\begin{equation}
S_{(m,n)} = \left(|V_m| \mathrm{e}^{\mathrm{j}\theta_m}\right) \kappa_{(m,n)}^\mathrm{H} \mathrm{diag}\left(\frac{\mathbb{1}_N}{|V|\mathrm{e}^{\mathrm{j}\theta}}\right) S,
\label{eq:Smn2}
\end{equation}
where we point out that with reference to the notation established in Section~\ref{sec:Notation}, 
\begin{equation*}
\mathrm{diag}\left(\frac{\mathbb{1}_N}{|V|\mathrm{e}^{\mathrm{j}\theta}}\right) = \mathrm{diag}\left(\frac{1}{|V_1| \mathrm{e}^{\mathrm j \theta_1}},\dots,\frac{1}{|V_n| \mathrm{e}^{\mathrm j \theta_N}}\right).
\end{equation*}
Since line power flows are often expressed as functions of angle differences, we find it useful to rewrite~\eqref{eq:Smn2} as
\begin{align*}
S_{(m,n)} &= |V_m| \kappa_{(m,n)}^\mathrm{H} \mathrm{diag}\left(\frac{\mathbb{1}_N \mathrm{e}^{\mathrm{j}\theta_m}}{|V|\mathrm{e}^{\mathrm{j}\theta}}\right) S \nonumber \\
&= |V_m|  \kappa_{(m,n)}^\mathrm{H} \mathrm{diag}\left(\frac{\mathrm{e}^{{\mathrm{j}\theta^m}}}{|V|}  \right) S ,
\end{align*}
where $\theta^m = \theta_m \mathbb{1}_{N} - \theta \in \mathbb{T}^N$.  To simplify the expression above, define:
\begin{align*}
\Xi_m +\mathrm j \Psi_m := \mathrm{diag} \left(\frac{\cos \theta^m}{| V|} \right) + \mathrm{j} \,\, \mathrm{diag} \left(\frac{\sin \theta^m}{| V|} \right).
\end{align*}
Making use of the above and the decomposition of the current injection sensitivity factors in~\eqref{eq:a}, we obtain
\begin{align*}
S_{(m,n)} &= |V_m|  \left(\alpha_{(m,n)}^\mathrm T - \mathrm j \beta_{(m,n)}^\mathrm T \right)  (\Xi_m + \mathrm j \Psi_m) (P + \mathrm j Q) \nonumber \\
&=|V_m|  \left(\alpha_{(m,n)}^\mathrm T \Xi_m + \beta_{(m,n)}^\mathrm T \Psi_m \right) (P + \mathrm j Q)  \nonumber \\
&+\mathrm j |V_m|  \left(\alpha_{(m,n)}^\mathrm T \Psi_m - \beta_{(m,n)}^\mathrm T \Xi_m \right) (P + \mathrm j Q).
\end{align*}
Then, we get the following \emph{power divider laws}  that indicate how bus active- and reactive-power  injections map to the active- and reactive-power flows on line $(m,n)$: 
\begin{align}
P_{(m,n)} &= \mathrm{Re}\{S_{(m,n)}\} =|V_m| \left(u_{(m,n)}^\mathrm{T} P - v_{(m,n)}^\mathrm{T} Q\right) \label{eq:Pmn}, \\
Q_{(m,n)} &= \mathrm{Im}\{S_{(m,n)}\}= |V_m| \left(u_{(m,n)}^\mathrm{T} Q + v_{(m,n)}^\mathrm{T} P\right) \label{eq:Qmn}.
\end{align}
In~\eqref{eq:Pmn}--\eqref{eq:Qmn}, $u_{(m,n)}, v_{(m,n)}  \in \mathbb{R}^{N}$ are given by 
\begin{align} \label{eq:u}
u_{(m,n)} &:= \Xi_m\alpha_{(m,n)} + \Psi_m\beta_{(m,n)} \\
					& = \mathrm{diag} \left(\frac{\cos \theta^m}{| V|} \right) \alpha_{(m,n)} + \mathrm{diag} \left(\frac{\sin \theta^m}{| V|} \right)\beta_{(m,n)}, \nonumber
\end{align}
\begin{align} \label{eq:v}
v_{(m,n)} &:= \Psi_m\alpha_{(m,n)} - \Xi_m\beta_{(m,n)} \\
& = \mathrm{diag} \left(\frac{\sin \theta^m}{| V|} \right) \alpha_{(m,n)} - \mathrm{diag} \left(\frac{\cos \theta^m}{| V|} \right)\beta_{(m,n)}. \nonumber
\end{align}
We note that~\eqref{eq:Pmn}--\eqref{eq:Qmn} are the complex-power analogues of the current divider in~\eqref{eq:amn_inv}, namely they indicate how the complex-power injection at each bus $i$ contributes to the complex-power flow on line~$(m,n)$.  We refer to~\eqref{eq:Pmn}--\eqref{eq:Qmn} as the power divider laws, since they specify how the active- and reactive-power flows on line~$(m,n)$ \emph{divide} among active- and reactive-power injections at each bus $i$.

\section{Approximations, Decoupling, and Connections to the DC Power Flow} \label{sec:approx}
In this section, we make use of a few practical observations and engineering insights regarding high-voltage transmission systems to present a suite of approximations to~\eqref{eq:Pmn}--\eqref{eq:Qmn} that are conceivably applicable in different contexts. We also uncover the classical DC power-flow expressions under the most restrictive set of assumptions.

\subsection{Lossless Networks} \label{sec:lossless}
The line resistance in transmission circuits is much lower than the line reactance~\cite{Glover:2012}.  As a direct consequence, the conductance is much smaller than the susceptance.  In other words, $g_{mn} << b_{mn}$ in the series admittance of line $(m,n) \in \mathcal{E}$ and $g_m << b_m$ in the shunt admittance at bus $m \in \mathcal{N}$.  Thus, we can approximate
$y_{mn} \approx \mathrm{j} b_{mn}$, $\ y_m \approx \mathrm{j} b_m$, and $\ Y \approx \mathrm{j} B$. With these approximations in place,~\eqref{eq:amn_inv} simplifies to the following:
\begin{equation} \label{eq:kappalossless}
\kappa_{(m,n)}^\mathrm{T} = \alpha_{(m,n)}^\mathrm{T} \approx  \left(  b_{mn} e_{mn}^\mathrm{T}  + b_m e_{m}^\mathrm{T}   \right) B^{-1}.
\end{equation}
Therefore, under the lossless assumption,~\eqref{eq:u}--\eqref{eq:v} become
\begin{align} 
u_{(m,n)} &\approx \Xi_m \alpha_{(m,n)} = \mathrm{diag}\left(\frac{\cos \theta^m}{|V|}\right)\alpha_{(m,n)} \label{eq:ulossless}, \\
v_{(m,n)} &\approx \Psi_m\alpha_{(m,n)} = \mathrm{diag}\left(\frac{\sin \theta^m}{|V|}\right)\alpha_{(m,n)}. \label{eq:vlossless}
\end{align}
The power divider expressions in~\eqref{eq:Pmn}--\eqref{eq:Qmn} are then given by
\begin{align}
P_{(m,n)} &=|V_m| \alpha_{(m,n)}^\mathrm{T} \left(\mathrm{diag}\left(\frac{\cos \theta^m}{|V|}\right) P \right. \nonumber \\
&\hspace{0.75in} -\left.\mathrm{diag}\left(\frac{\sin \theta^m}{|V|}\right) Q\right), \label{eq:Pmn_lossless} \\
Q_{(m,n)} &=|V_m| \alpha_{(m,n)}^\mathrm{T} \left(\mathrm{diag}\left(\frac{\cos \theta^m}{|V|}\right) Q \right. \nonumber \\&\hspace{0.75in} +\left. \mathrm{diag}\left(\frac{\sin \theta^m}{|V|}\right) P\right). \label{eq:Qmn_lossless}
\end{align}

\subsection{Small Angle Differences}
Under typical operating conditions, the phase angle differences between any two transmission-system buses are small, i.e., $\theta_i - \theta_k << 1$, for all $i, k \in \mathcal{N}$.  It follows that $\sin(\theta_i - \theta_k) \approx \theta_i - \theta_k$ and $\cos(\theta_i - \theta_k) \approx 1$.  Under these assumptions,~\eqref{eq:ulossless}--\eqref{eq:vlossless} further simplify to
\begin{align}
u_{(m,n)}  &\approx \mathrm{diag} \left(\frac{\mathbb{1}_N}{| V|} \right) \alpha_{(m,n)},  \label{eq:u_smallangle}\\
v_{(m,n)} &\approx \mathrm{diag} \left(\frac{\theta^m}{| V|} \right)  \alpha_{(m,n)}. \label{eq:v_smallangle}
\end{align}
The power divider laws in~\eqref{eq:Pmn_lossless}--\eqref{eq:Qmn_lossless} are then given by
\begin{align}
P_{(m,n)} &=|V_m| \alpha_{(m,n)}^\mathrm{T} \left(\mathrm{diag}\left(\frac{\mathbb{1}_N}{|V|}\right) P - \mathrm{diag}\left(\frac{\theta^m}{|V|}\right) Q\right), \label{eq:Pmn_smallangle} \\
Q_{(m,n)} &=|V_m| \alpha_{(m,n)}^\mathrm{T} \left(\mathrm{diag}\left(\frac{\mathbb{1}_N}{|V|}\right) Q + \mathrm{diag}\left(\frac{\theta^m}{|V|}\right) P\right). \label{eq:Qmn_smallangle}
\end{align}

\subsection{Unity Voltage Magnitude} \label{sec:UnitVoltage}
In the per-unit system, the numerical values of voltage magnitudes are typically such that $|V| \approx \mathbb{1}_N$.  Thus,~\eqref{eq:u_smallangle}--\eqref{eq:v_smallangle} simplify to
\begin{align}
u_{(m,n)}  &\approx \alpha_{(m,n)},  \label{eq:u_unit}\\
v_{(m,n)} &\approx \mathrm{diag} \left(\theta^m \right)  \alpha_{(m,n)}. \label{eq:v_unit}
\end{align}
Accordingly, the approximate power divider expressions in~\eqref{eq:Pmn_smallangle}--\eqref{eq:Qmn_smallangle} are then given by
\begin{align}
P_{(m,n)} &= \alpha_{(m,n)}^\mathrm{T} \left(P - \mathrm{diag}\left(\theta^m\right) Q\right), \label{eq:Pmn_unity} \\
Q_{(m,n)} &= \alpha_{(m,n)}^\mathrm{T} \left(Q + \mathrm{diag}\left(\theta^m\right) P\right). \label{eq:Qmn_unity}
\end{align}

\subsection{Decoupled Power Flow} \label{sec:decouple}
Since angle differences are expectedly small, it follows that $e_i^\mathrm T\theta^m Q_i<< P_i$ and $e_i^\mathrm T\theta^m P_i<< Q_i$, for all $i \in \mathcal N$. With this assumption in place, the expressions in~\eqref{eq:Pmn_unity}--\eqref{eq:Qmn_unity} further simplify to
\begin{align}
P_{(m,n)} &= \alpha_{(m,n)}^\mathrm{T} P, \label{eq:Pmn_decouple} \\
Q_{(m,n)} &= \alpha_{(m,n)}^\mathrm{T} Q. \label{eq:Qmn_decouple}
\end{align}
The above expressions echo the widely accepted decoupling assumptions between active- and reactive-power in transmission systems.  Note that~\eqref{eq:Pmn_decouple}--\eqref{eq:Qmn_decouple} are valid only when the entries of $P$ and $Q$ are of the same order of magnitude. For instance, if reactive-power injections are close to zero then the approximation in~\eqref{eq:Qmn_decouple} would no longer be valid.  

\subsection{Connection to the DC Power Flow} \label{rem:DCPowerFlow}
Here, we demonstrate that the DC power-flow equations can be derived from the active-power flow expression in~\eqref{eq:Pmn_decouple} under the further approximation of neglecting shunt susceptance terms in a lossless transmission network.  In order to uncover this relationship, a slack bus needs to be assigned (since the construction of the DC power flow equations requires a slack bus).  Without loss of generality, set bus 1 as the slack bus.  Accordingly, partition $\alpha_{(m,n)}^\mathrm T = [\alpha_{(m,n),1}, \widetilde\alpha_{(m,n)}^\mathrm T]$, where $\alpha_{(m,n),1} = e_1^\mathrm T \alpha_{(m,n)} \in \mathbb{R}$ isolates the element of $\alpha_{(m,n)}$ that corresponds to the slack bus, and $\widetilde\alpha_{(m,n)} = \Pi_1^\mathrm T \alpha_{(m,n)} \in \mathbb{R}^{N-1}$ collects the remaining elements.  Similarly, partition $P = [P_1, \widetilde P^1]$, where $\widetilde P^1 = \Pi_1^\mathrm T P$.  With these partitioned vectors,~\eqref{eq:Pmn_decouple} becomes
\begin{align}
P_{(m,n)} &=  \alpha_{(m,n),1} P_1 + \widetilde\alpha_{(m,n)}^\mathrm T \widetilde P^1 \nonumber \\
& = \left( \widetilde\alpha_{(m,n)}^\mathrm T - \alpha_{(m,n),1} \mathbb{1}_{N-1}^\mathrm T \right) \widetilde P^1, \label{eq:Pmn_partition}
\end{align}
where the second equality above results from substituting $P_1 = -\mathbb{1}_{N-1}^\mathrm T \widetilde P^1$, which is valid for lossless systems, since $\mathbb 1_N^\mathrm T P = P_1 + \mathbb 1_{N-1}^\mathrm T \widetilde P^1 = 0$. 

\begin{figure}[t!]
\centering
\mbox{
\epsfig{file=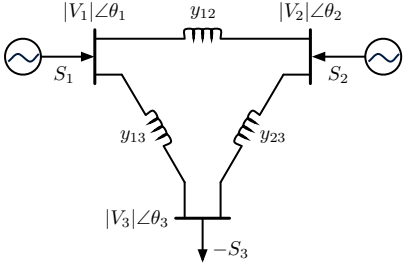,width=0.3\textwidth}}
\caption{Network topology for 3-bus system.}
\label{fig:3bus}
\vspace{-7pt}
\end{figure}

Next, we examine the structure of $\alpha_{(m,n)}^\mathrm T$ for a system in which losses and shunt elements are omitted.  Neglecting shunt susceptance terms (i.e., setting $b_m = 0,\,\, \forall\, m \in \mathcal N$) in~\eqref{eq:kappalossless}, we obtain $\alpha_{(m,n)}^\mathrm T  = b_{mn} e_{mn}^{\mathrm T} B^{\dagger}$.  Note that the bus susceptance matrix $B$ is not invertible since ground-connecting shunt elements are removed, so we resort to using the pseudo-inverse of $B$, namely $B^\dagger$ (see Remark~\ref{rem:Yinv}).  We can equivalently express $\alpha_{(m,n)}^\mathrm T B = b_{mn} e_{mn}^{\mathrm T}$.  Partitioning out bus~1 (the slack bus) in $\alpha_{(m,n)}$, $B$, and $e_{mn}$, we get that
\begin{equation}
\begin{bmatrix} \alpha_{(m,n),1} & \widetilde \alpha_{(m,n)}^\mathrm T \end{bmatrix} \begin{bmatrix} b_{1,1} & \widetilde b_1^\mathrm T \\ \widetilde b_1 & \widetilde B \end{bmatrix} = b_{mn} \begin{bmatrix} e_{mn,1} & \widetilde{e}_{mn}^\mathrm T\end{bmatrix},
\label{eq:alpha_partition}
\end{equation}
where $\widetilde B$ is constructed by removing the row and column corresponding to the slack bus from $B$. Focusing on the second element in~\eqref{eq:alpha_partition}, we obtain
\begin{equation}
\alpha_{(m,n),1} \widetilde b_1^\mathrm T + \widetilde \alpha_{(m,n)}^\mathrm T \widetilde B = b_{mn} \widetilde{e}_{mn}^\mathrm T.
\label{eq:alpha_partition2}
\end{equation}
Furthermore, since shunt elements are neglected, $B$ satisfies $B \mathbb{1}_N = \mathbb{0}_N$, from which we get $\widetilde{b}_1 = - \widetilde B \mathbb{1}_{N-1}$.  By substituting this into~\eqref{eq:alpha_partition2}, it simplifies to
\begin{align}
\left( \widetilde \alpha_{(m,n)}^\mathrm T - \alpha_{(m,n),1} \mathbb{1}_{N-1}^\mathrm T \right) \widetilde B &= b_{mn} \widetilde{e}_{mn}^\mathrm T \nonumber \\
 \widetilde \alpha_{(m,n)}^\mathrm T - \alpha_{(m,n),1} \mathbb{1}_{N-1}^\mathrm T &= b_{mn} \widetilde{e}_{mn}^\mathrm T \widetilde B^{-1} .
\label{eq:alpha_part}
\end{align}
where the second equality above results from recognizing that the reduced bus susceptance matrix $\widetilde B$ is invertible.  Substituting~\eqref{eq:alpha_part} into~\eqref{eq:Pmn_partition}, we  obtain 
\begin{equation}
P_{(m,n)} = b_{mn} \widetilde{e}_{mn}^\mathrm T \widetilde B^{-1} \widetilde P^1.
\label{eq:Pmn_dc}
\end{equation}  
The classical expression for the power flow on a line with a small-angle approximation is given by~\cite{Glover:2012} 
\begin{equation}
P_{(m,n)} = -b_{mn} (\theta_m - \theta_n) = -b_{mn} \widetilde e_{mn}^\mathrm T \widetilde \theta^1, 
\label{eq:Pmn_dc2}
\end{equation}
where $\widetilde \theta^1$ is as defined in~\eqref{eq:theta_defs}.  Finally, by equating~\eqref{eq:Pmn_dc2} with~\eqref{eq:Pmn_dc}, we get $\widetilde \theta^1 =  - \widetilde B^{-1} \widetilde P^1$, which are the DC power-flow expressions~\cite{Glover:2012}.

\begin{example}[Approximations]
\label{ex:3bus1}
Consider the 3-bus system with the one-line diagram shown in Fig.~\ref{fig:3bus}.  Transmission lines are modelled with lumped parameters, where $y_{12} = 1.3652 - \mathrm{j} 11.6041\, \mathrm{p.u.}$, $y_{12}^\mathrm{sh} = \mathrm{j}0.088\, \mathrm{p.u.}$, $y_{23} = 0.7598 - \mathrm{j} 6.1168\, \mathrm{p.u.}$, $y_{23}^\mathrm{sh} = \mathrm{j}0.153\, \mathrm{p.u.}$, $y_{13} = 1.1677 - \mathrm{j} 10.7426\, \mathrm{p.u.}$, $y_{13}^\mathrm{sh} = \mathrm{j}0.079\, \mathrm{p.u.}$   In this system, generators are connected to buses 1 and 2, injecting $P_1 = 1.5973\, \mathrm{p.u.}$ and $P_2 = 0.7910\, \mathrm{p.u.}$, respectively.  A $PQ$ load is connected to bus 3, with $P_3 = -2.35\, \mathrm{p.u.}$ and $Q_3 = -0.5\, \mathrm{p.u.}$  The voltage magnitudes at buses 1 and 2 are regulated at $|V_1| = 1.04\, \mathrm{p.u.}$ and $|V_2| = 1.025\, \mathrm{p.u.}$, respectively.  Using the parameters listed above, we compute the power flow solution and then proceed to compute the line active- and reactive-power flows using the exact expressions in~\eqref{eq:Pmn} and~\eqref{eq:Qmn}, respectively.  Subsequently, we approximate the line flows using the simplifying assumptions outlined in Section~\ref{sec:approx}, and results are collected in Table~\ref{table:3bus_approx}. \qedblack 
\end{example} 

\begin{table}[t!]
\begin{center}
\begin{scriptsize}
\caption{Exact and approximate line active- and reactive-power flows for Example~\ref{ex:3bus1}.  All quantities are in $\mathrm{p.u.}$}
\begin{tabular}{@{} c | c  c c c c}
Line & Exact & Lossless & Small  & Unity       & DC \\
Flow   &    &         & Angle &  Magnitude  & Power Flow \\
& \eqref{eq:Pmn}--\eqref{eq:Qmn}   &     \eqref{eq:Pmn_lossless}--\eqref{eq:Qmn_lossless}       & \eqref{eq:Pmn_smallangle}--\eqref{eq:Qmn_smallangle} &   \eqref{eq:Pmn_unity}--\eqref{eq:Qmn_unity}  & \eqref{eq:Pmn_dc2} \\
\hline
$P_{(1,2)}$ & $0.0533$  & $0.0515$  & $0.0461$  & $0.0753$  & $0.0300$\\
 $P_{(2,3)}$ & $0.844$  & $0.843$  & $0.843$  & $0.847$  & $0.800$      \\
$P_{(1,3)}$ & $1.54$  & $1.55$  & $1.55$  & $1.52$  & $1.43$  \\
\hline
$Q_{(1,2)}$ & $0.0821$  & $0.0894$  & $0.0880$  & $0.0965$  & --- \\
$Q_{(2,3)}$ & $-0.0123$  & $-0.0061$  & $-0.0059$  & $-0.0051$  & ---      \\
$Q_{(1,3)}$ & $0.370$  & $0.363$  & $0.364$  & $0.356$  & ---  \\
\end{tabular}
\label{table:3bus_approx}
\end{scriptsize}
\end{center}
\vspace{-10pt}
\end{table}

\section{Applications} \label{sec:app}

In this section, we detail three applications for the power divider laws derived in Section~\ref{sec:line} and their approximations described in Section~\ref{sec:approx}. 

\subsection{Transmission-network Allocation}
With the power divider laws, we can explicitly uncover the proportion of bus injections that contribute to line flows in analytical closed form. In the literature, this is typically referred to as transmission-network allocation, and when electricity prices are factored in, as transmission-network cost allocation~\cite{Conejo-2007}  From~\eqref{eq:Pmn} and~\eqref{eq:Qmn}, we can express the active- and reactive-power flows on line $(m,n)$ as the following sum of $2N$ scalar terms:
\begin{align}
P_{(m,n)} &= |V_m| \left(\sum_{i = 1}^N  e_i^\mathrm T u_{(m,n)} P_i -  \sum_{i = 1}^N e_i^\mathrm T v_{(m,n)} Q_i \right), \label{eq:Pmn_sum} \\
Q_{(m,n)} &= |V_m| \left( \sum_{i = 1}^N   e_i^\mathrm T u_{(m,n)} Q_i + \sum_{i = 1}^N   e_i^\mathrm T v_{(m,n)} P_i \right). \label{eq:Qmn_sum}
\end{align}
With~\eqref{eq:Pmn_sum}, for each bus $i$, we compute the contributions of its active- and reactive-power injection components to the active-power flow in line $(m,n)$ respectively as
\begin{equation}
\frac{|V_m| e_i^\mathrm T u_{(m,n)} P_i}{P_{(m,n)}}, \quad - \frac{|V_m| e_i^\mathrm T v_{(m,n)} Q_i}{P_{(m,n)}}.
\label{eq:Pmn_contrib}
\end{equation}
Similarly, from~\eqref{eq:Qmn_sum}, we compute the contributions of the active- and reactive-power injection components at bus $i$ to the reactive-power flow in line $(m,n)$ respectively as
\begin{equation}
\frac{|V_m| e_i^\mathrm T v_{(m,n)} P_i}{Q_{(m,n)}}, \quad \frac{|V_m| e_i^\mathrm T u_{(m,n)} Q_i}{Q_{(m,n)}}.
\label{eq:Qmn_contrib}
\end{equation}

\begin{example}[Active-power Transmission Allocation]
\label{ex:3bus2}
Consider the 3-bus system from Example~\ref{ex:3bus1}. Without loss of generality, let us focus on the active-power flow on line $(1,3)$, which is $P_{(1,3)} = 1.5440\,\mathrm{p.u.}$ Computed using~\eqref{eq:Pmn_contrib}, the bus active-power injection contributions to $P_{(1,3)}$ from buses 1, 2, and 3 are $49.88\%$, $12.11\%$, and $39.19\%$, respectively.  The bus reactive-power injection contributions are on the order of $1\%$. These quantities can be directly used by power system operators to apportion the cost of transmitting active power on lines in the network to generators and loads.
\qedblack
\end{example}

\subsection{Transmission-loss Allocation}
\label{sec:loss}
The expression in~\eqref{eq:Pmn} can be leveraged to uncover the contributions of bus active- and reactive-power injections to active-power transmission losses on each line. In the literature, this general task is typically termed as transmission-loss allocation~\cite{Conejo-2001}. Denote the active-power loss on line $(m,n)$ as $L_{(m,n)}$, and consider that $L_{(m,n)}$ can be expressed as 
\begin{align}
L_{(m,n)} &= |I_{(m,n)} - y_m V_m|^2 \mathrm{Re}\{y_{mn}^{-1}\} \label{eq:I2R1} \\
&= \mathrm{Re}\left\{y_{mn}^{-1} (I_{(m,n)} - y_m V_m)^*(I_{(m,n)} - y_m V_m)\right\}. \nonumber
\end{align}
Recognizing from~\eqref{eq:Ikl_mtx} that $I_{(m,n)} - y_m V_m = (V_m - V_n) y_{mn}$, and substituting this into~\eqref{eq:I2R1}, we obtain
\begin{equation}
L_{(m,n)} =  \mathrm{Re}\left\{(V_m - V_n) y_{mn}^* (V_m - V_n)^*\right\}.
\label{eq:I2R2}
\end{equation}
Further suppose the shunt elements are purely imaginary, i.e., $y_m = \mathrm{j} b_m$, which is a common assumption in transmission line modelling, then~\eqref{eq:I2R2} can be equivalently expressed as
\begin{align}
L_{(m,n)} &=  \mathrm{Re}\left\{(V_m - V_n) y_{mn}^* (V_m - V_n)^*\right. \nonumber \\
    & \quad + \left. V_m y_m^* V_m^* + V_n y_n^* V_n^* \right\}, \label{eq:I2R3}
\end{align}
where the addition of $V_m y_m^* V_m^* + V_n y_n^* V_n^*$, does not affect the real part of the expression. Finally, with $I_{(m,n)}$ in~\eqref{eq:Ikl}, we simplify~\eqref{eq:I2R3} as
\begin{equation*}
L_{(m,n)} =  \mathrm{Re}\left\{ V_m I_{(m,n)}^* + V_n I_{(n,m)}^* \right\} = P_{(m,n)} + P_{(n,m)},
\end{equation*}
where $P_{(m,n)}$ ($P_{(n,m)}$) represents the active-power flow from bus $m$ ($n$) to $n$ ($m$) along line $(m,n)$. With the aid of~\eqref{eq:Pmn}, we can write
\begin{align}
L_{(m,n)} &= \left( |V_m| u_{(m,n)}^\mathrm{T}  +  |V_n| u_{(n,m)}^\mathrm{T} \right) P \nonumber \\
& +   \left( |V_m| v_{(m,n)}^\mathrm{T}  +  |V_n| v_{(n,m)}^\mathrm{T} \right) Q,
\label{eq:loss}
\end{align}
which indicates how bus active- and reactive-power injections contribute to the active-power line loss on the $(m,n)$ line.

Corresponding to each one of the cases in Section~\ref{sec:approx}, the expression for the transmission-network losses in~\eqref{eq:loss} also simplifies correspondingly. In the particular case considered in Section~\ref{sec:decouple}, the expression for the losses on the $(m,n)$ line simplifies to $L_{(m,n)} = (\alpha_{(m,n)} + \alpha_{(n,m)}) P$, which resonates with prior art that has used current flows/injections as proxies for power flows/injections~\cite{Conejo-2007}.

\begin{example}[Transmission-line Losses]
\label{ex:3bus3}
Consider, again, the 3-bus system from Example~\ref{ex:3bus1}.  To demonstrate the concepts above, using~\eqref{eq:loss}, transmission-line losses are obtained for each line as $L_{(1,2)} = 0.0003\,\mathrm{p.u.}$, $L_{(2,3)} = 0.0140\,\mathrm{p.u.}$, and $L_{(1,3)} = 0.0240\,\mathrm{p.u.}$  The total system loss can be computed by summing up losses in all lines as $0.0383\,\mathrm{p.u.}$  We verify this quantity by summing up all bus active-power injections, specifically, $P_1 = 1.60\,\mathrm{p.u.}$, $P_2 = 0.791\,\mathrm{p.u.}$, and $P_3 = -2.35\,\mathrm{p.u.}$  \qedblack
\end{example}

\subsection{Line Active-power Flow-constrained Injections} 
Here, we consider the problem of obtaining a set of bus active-power injections that best satisfy a set of desired transmission-line active-power flows. In order to achieve this goal, below, we formulate a convex optimization problem that utilizes the power divider approximation in Section~\ref{sec:decouple}.

Let $\mathcal{D} \subseteq \mathcal{E}$ denote the subset of $D$ transmission lines for which active-power flows are specified and suppose $D \geq N$. Collect these $D$ designated reference line flows, i.e., $\{P_{(m,n)}^\mathrm r\}$, $(m,n) \in \mathcal{D}$  into the column vector $P_{\mathcal{D}} \in \mathbb{R}^{D}$ and corresponding real components of current injection sensitivity factors, i.e., $\{\alpha_{(m,n)}\}$,  $(m,n) \in \mathcal{D}$, into the matrix $A \in \mathbb{R}^{D \times N}$.  In order to obtain a set of bus active-power injections that best satisfies the desired line flows and simultaneously adheres to system power balance, we solve for $P\in \mathbb{R}^N$ from the following linearly constrained least-squares optimization problem~\cite{Boyd:2004}: 
\begin{align}
\min_{P \in \mathbb{R}^N} & \quad ||A P - P_{\mathcal{D}}||^2 \nonumber \\
\mathrm{s.t.} & \quad \mathbb{1}_N^\mathrm T P = L,
\label{eq:lse}
\end{align}
where $L$ denotes the total system loss. The unique closed-form solution to~\eqref{eq:lse} is given by:
\begin{equation}
\begin{bmatrix} P \\ \lambda \end{bmatrix} = \begin{bmatrix} 2A^\mathrm{T} A & \mathbb{1}_N \\ \mathbb{1}_N^{\mathrm{T}} & 0 \end{bmatrix}^{-1} \begin{bmatrix} 2A^\mathrm{T} P_{\mathcal{D}} \\ L \end{bmatrix},
\label{eq:lse_sol}
\end{equation}
where $\lambda$ is the Lagrange multiplier associated with the equality constraint. Invertibility of the matrix in~\eqref{eq:lse_sol} can be ensured \emph{if and only if} $[A^\mathrm{T}, \mathbb{1}_N]^\mathrm{T}$ has independent columns~\cite{Vandenberghe:2015}. Note that the total system loss $L$ in~\eqref{eq:lse} is not known \textit{a priori}. One option is to set it to zero, as would approximately be the case in transmission networks with dominantly inductive lines. An alternative approximation is highlighted next. 

Neglecting shunt elements in~\eqref{eq:I2R1}, we get that $L_{(m,n)} = |I_{(m,n)}|^2 \mathrm{Re}\{ y_{mn}^{-1}\}$.  Recall, from Section~\ref{sec:loss}, that current flows/injections can act as proxies for power flows/injections.  Consequently, we can approximate $L_{(m,n)} \approx |S_{(m,n)}|^2 \mathrm{Re}\{ y_{mn}^{-1}\}$.  Furthermore, suppose that the power factor on each line $(m,n)$ is close to unity, then we can express the expected loss on line $(m,n)$ as $L_{(m,n)}^\mathrm r = |P_{(m,n)}^\mathrm r|^2 \mathrm{Re}\{ y_{mn}^{-1}\}$, where $P_{(m,n)}^\mathrm r$ denotes the desired active-power flow on line $(m,n)$.  Similar to the construction of $P_\mathcal{D}$, collect these predicted line losses, i.e., $\{L_{(m,n)}^\mathrm r\}$, $(m,n) \in \mathcal{D}$, into the column vector $L_\mathcal{D} \in \mathbb{R}^{D}$.  With these variables in place, we reformulate~\eqref{eq:lse} with $L = \mathbb{1}_{D}^\mathrm T L_\mathcal{D}$, resulting in the following unique closed-form solution:
\begin{equation}
\begin{bmatrix} P \\ \lambda \end{bmatrix} = \begin{bmatrix} 2A^\mathrm{T} A & \mathbb{1}_N \\ \mathbb{1}_N^{\mathrm{T}} & 0 \end{bmatrix}^{-1} \begin{bmatrix} 2A^\mathrm{T} P_{\mathcal{D}} \\  \mathbb{1}_{D}^\mathrm T L_\mathcal{D} \end{bmatrix}.
\label{eq:lselossy_sol}
\end{equation}

\begin{figure*}[t!]
        \centering
        \begin{subfigure}[b]{0.49\textwidth}
                \centering
                \includegraphics[width=1\textwidth]{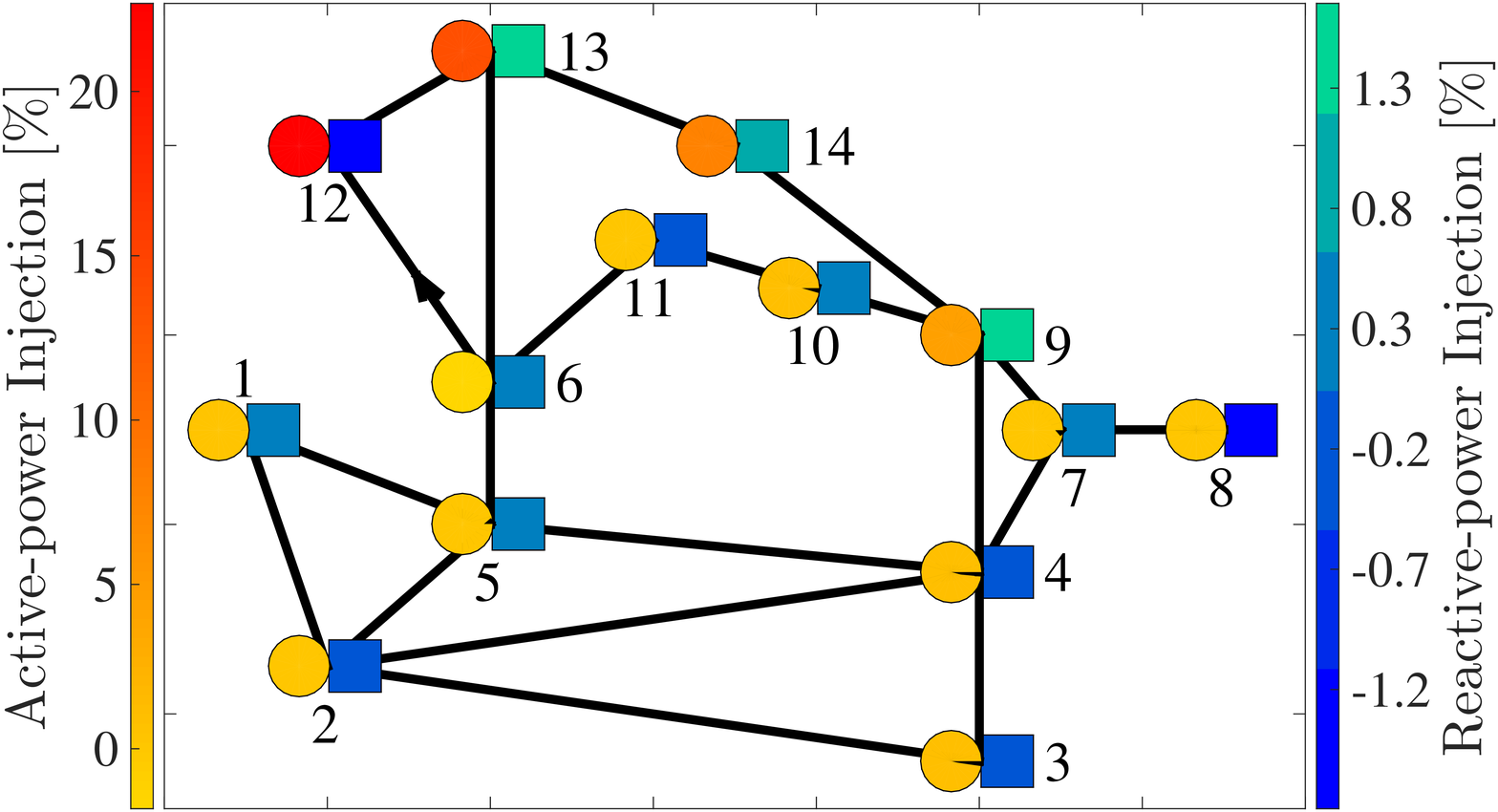}
                \caption{Active-power flow.}
                \label{fig:line7Pflow}
        \end{subfigure}%
        ~~
        \begin{subfigure}[b]{0.49\textwidth}
                \centering
                \includegraphics[width=1\textwidth]{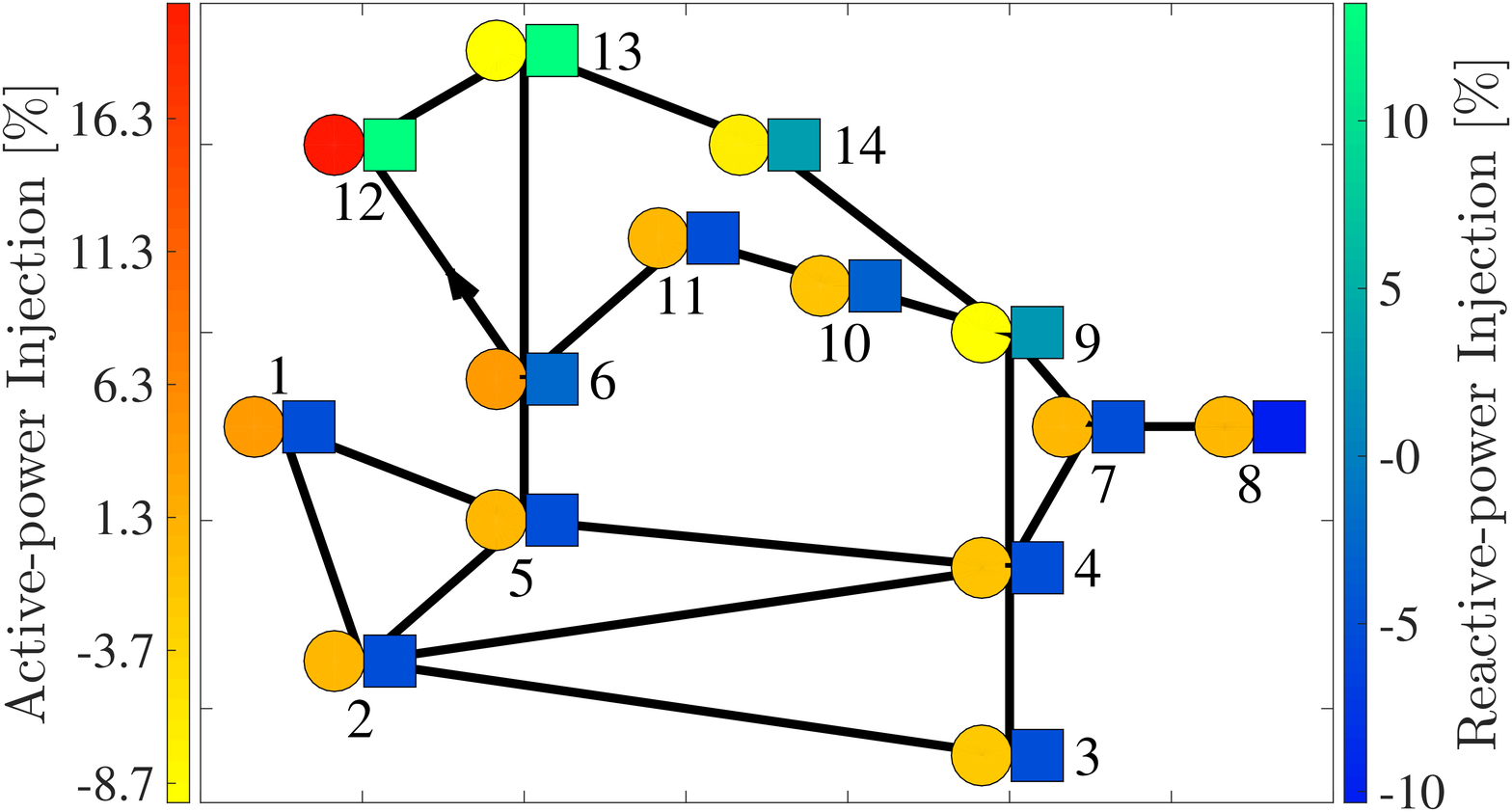}
                \caption{Reactive-power flow.}
                \label{fig:line7Qflow}
        \end{subfigure} 
        \caption{IEEE 14-bus system: Contribution of bus active- and reactive power (marked as circle and square, respectively) injections to active- and reactive-power flow in line $(6,12)$ (arrow indicates direction of flow).  Each bus is associated with a coloured circle and a coloured square, with the colour representative of the bus' contribution to the power flow in line $(6,12)$.}
        \label{fig:14bus_divider}
        \vspace{-10pt}
\end{figure*}

\begin{example}[Constrained Injections]
\label{ex:3bus5}
Consider the 3-bus system from Example~\ref{ex:3bus1}.  Using~\eqref{eq:amn_inv} and~\eqref{eq:a}, we get $\alpha_{(1,2)} = [0.518, -0.233, 0.249]^\mathrm T$, $\alpha_{(2,3)} = [ 0.244, 0.493, -0.0289]^\mathrm T$, and $\alpha_{(1,3)} = [ 0.482, 0.233, -0.249]^\mathrm T$.  Suppose the desired reference active-power flows on lines in the set $\mathcal{D} = \{(1,2), (2,3), (1,3)\}$ are $P_{(1,2)}^{\mathrm{r}} = 0.46\,\mathrm{p.u.}$, $P_{(2,3)}^{\mathrm{r}} = 0.67\,\mathrm{p.u.}$, and  $P_{(1,3)}^{\mathrm{r}} = 1.65\,\mathrm{p.u.}$.   We have thereby assembled all ingredients of the optimization problem in~\eqref{eq:lse}, where $A = [\alpha_{(1,2)}, \alpha_{(2,3)}, \alpha_{(1,3)}]^\mathrm T$, and $P_\mathcal{D} = [P_{(1,2)}^{\mathrm{r}}, P_{(2,3)}^{\mathrm{r}}, P_{(1,3)}^{\mathrm{r}}]^\mathrm T$.

Based on desired line active-power flows above, the expected line losses are collected in $L_\mathcal{D} = [L_{(1,2)}^\mathrm r, L_{(2,3)}^\mathrm r, L_{(1,3)}^\mathrm r]^\mathrm T$, where $L_{(1,2)}^\mathrm r = 0.0021\,\mathrm{p.u.}$, $L_{(2,3)}^\mathrm r = 0.0090\,\mathrm{p.u.}$, and $L_{(1,3)}^\mathrm r = 0.0272\,\mathrm{p.u.}$  Via~\eqref{eq:lselossy_sol}, we get that $P_1 = 2.11\,\mathrm{p.u.}$, $P_2 = 0.222\,\mathrm{p.u.}$, and $P_3 = -2.29\,\mathrm{p.u.}$  With these bus active-power injections, the nonlinear power flow solution reveals that the actual line active-power flows are $P_{(1,2)} = 0.468\,\mathrm{p.u.}$, $P_{(2,3)} = 0.688\,\mathrm{p.u.}$, and $P_{(1,3)} = 1.64\,\mathrm{p.u.}$  
The 2-norm of the deviation of the actual line active-power flows from the desired ones is $0.0218\,\mathrm{p.u.}$  
The actual system total loss is $0.0384\,\mathrm{p.u.}$, which is approximately equal to the expected value of $\mathbb{1}_{D}^\mathrm T L_\mathcal{D} =  0.0383\,\mathrm{p.u.}$

Next, neglecting losses in~\eqref{eq:lse_sol} with $L = 0$, we get $P_1 = 2.11\,\mathrm{p.u.}$, $P_2 = 0.208\,\mathrm{p.u.}$, and $P_3 = -2.32\,\mathrm{p.u.}$  With these updated bus active-power injections, we compute the nonlinear power flow solution and find that the actual line active-power flows are $P_{(1,2)} = 0.486\,\mathrm{p.u.}$, $P_{(2,3)} = 0.692\,\mathrm{p.u.}$, and $P_{(1,3)} = 1.66\,\mathrm{p.u.}$  The 2-norm of the deviation of the actual line active-power flows from the desired ones is $0.0360\,\mathrm{p.u.}$  Higher errors are attributable to neglecting losses. \qedblack
\end{example}

\section{Case Studies} \label{sec:cases}
In this section, we illustrate concepts presented in Sections~\ref{sec:line}--\ref{sec:app} using the IEEE 14-bus system.  The simulation tool MATPOWER~\cite{Zimmerman:2011} is used throughout to compute  all power-flow solutions.

\subsection{Transmission-network Allocation}
In this case study, as depicted in Fig.~\ref{fig:14bus_divider}, we visualize the proportion of bus active- and  reactive-power injections that contribute to active- and reactive-power flows in line $(6,12)$.  In Fig.~\ref{fig:line7Pflow}, the circle at each bus $i$ represents the percent contribution of its active-power component, i.e., the $i$th term in $|V_6| u_{(6,12)}^\mathrm T P$, to the active-power flow in line $(6,12)$, and the square represents the percent contribution of its reactive-power component, i.e., the $i$-th term in $-|V_6| v_{(6,12)}^\mathrm T Q$, both computed via~\eqref{eq:Pmn_contrib}.  Similarly, in Fig.~\ref{fig:line7Qflow}, the circles and squares indicate the bus active- and reactive-power injection contributions, respectively, to the reactive-power flow on line $(6,12)$, computed via~\eqref{eq:Qmn_contrib}.  As described in Section~\ref{sec:decouple}, we note that, indeed, there exists strong coupling between the line active-power flow and bus active-power injections.  On the other hand, the coupling is not evident between the line reactive-power flow and bus reactive-power injections.  As mentioned in Section~\ref{sec:decouple}, this is because the power factors of complex-power injections  at most buses are close to unity.

\subsection{Transmission-loss Allocation}

\begin{figure}[t!]
\centering
\mbox{
\epsfig{file=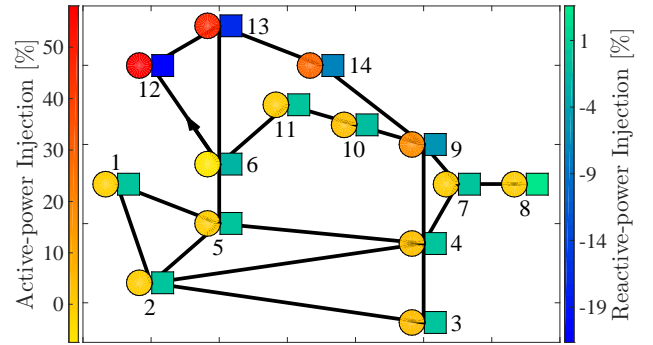,width=0.48\textwidth}}
\caption{IEEE 14-bus system: Contribution of bus active and reactive power (marked as circle and square, respectively) injections to active-power loss on line $(6,12)$.}
\label{fig:line7loss}
\vspace{-10pt}
\end{figure}

In this case study, we uncover the proportion of bus active- and reactive-power injections that contribute to active-power loss in line $(m,n)\in \mathcal{E}$.  We compute transmission losses for each line $(m,n)$ in the IEEE 14-bus system via the expression in~\eqref{eq:loss}.   Similar to~\eqref{eq:Pmn_sum}, we can rewrite~\eqref{eq:loss} as the sum of $2N$ terms and express, for each bus $i$, the contribution of its active-power injection to the loss on line $(m,n)$ as
\begin{equation}
\frac{e_i^T \left( |V_m| u_{(m,n)}  +  |V_n| u_{(n,m)} \right) P_i}{L_{(m,n)}}.
\label{eq:Lmn_contribP}
\end{equation}
Analogously, for each bus $i$, the contribution of its reactive-power injection to loss on line $(m,n)$ can be expressed as
\begin{equation}
\frac{e_i^T \left( |V_m| v_{(m,n)}  +  |V_n| v_{(n,m)} \right) Q_i}{L_{(m,n)}}.
\label{eq:Lmn_contribQ}
\end{equation}
In Fig.~\ref{fig:line7loss}, we plot the bus active- and reactive-power injection contributions to the active-power loss on line $(6,12)$.  Note that, unlike for the line active-power flow, there is no clear decoupling between active-power loss in line $(m,n)$ and bus reactive-power injections.  For example, $27.4\%$ of the loss in line $(6,12)$ is contributed from the active-power injection at bus $14$ and $-16.8\%$ is from the reactive-power injection at bus $13$.  The negative sign is indicative of injection directionality (i.e., generation versus load).

\subsection{Line Active-power Flow-constrained Injections} 
\begin{figure}[t!]
\centering
\mbox{
\epsfig{file=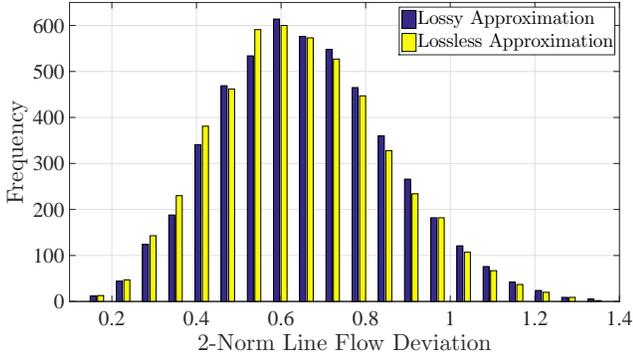,width=0.48\textwidth}}
\caption{IEEE 14-bus system: Histogram of 5000 line active-power flow-constrained injection simulations. The horizontal axis represents the 2-norm of the error between the actual line active-power flows and the desired ones, while the vertical axis represents the frequency with which a particular deviation range occurs.}
\label{fig:loadability}
%\vspace{-10pt}
\end{figure}

In this case study, using the IEEE 14-bus test system, we illustrate the application of the power divider laws to  obtain a set of active-power injections that best satisfy a set of desired transmission-line active-power flows.  To simulate these desired reference line flow values, for each line $(m,n) \in \mathcal{E}$, we apply random perturbations to the base-case transmission-line flows:
\begin{equation}
P_{(m,n)}^\mathrm r = P_{(m,n)} \left( 1 + \sigma \right),
\end{equation}
where $\sigma$ is a random variable uniformly distributed between $-1$ and $+1$.  With these randomly generated desired flow values, both lossy and lossless assumptions are applied and corresponding solutions are obtained via~\eqref{eq:lselossy_sol} and~\eqref{eq:lse_sol} with $L = 0$, respectively.  For each set of resulting bus active-power injections, the nonlinear power flow solution is computed to obtain the actual line active-power flows.  These are then compared to the desired reference values, and the 2-norms of errors between the actual flows and the desired ones are recorded.  In total, we run 5000 simulation cases and plot the histogram of errors in Fig.~\ref{fig:loadability}.  We note that the lossless approximation leads to line flows that are closer to the desired reference values.  This is likely due to the fact that the magnitude of the complex-power flows cannot be approximated by their real components only, i.e., the power factors of actual flows are not sufficiently close to unity.  Finally, we mention that it is quite likely that the randomly generated desired set of line active-power flows is, in fact, infeasible, and hence may lead to actual line flows with large deviations from the desired ones.

\section{Concluding Remarks and Directions for Future Work}
\label{sec:conc}

In this paper, we derive the so-called power divider laws, which indicate how active- and reactive-power flows on a transmission line are divided among bus active- and reactive-power injections.  A suite of approximations are then derived based on simplifying assumptions commonly applied to the analysis of transmission networks. We present three applications of the power divider laws: transmission-network allocation, transmission-loss allocation, and line flow-constrained injection estimation. Compelling avenues for future work are to leverage power divider laws for applications such as spot pricing, transmission-services pricing, and visualization.
\balance

\bibliographystyle{IEEEtran}
\bibliography{CC_bibliography}

%\begin{IEEEbiographynophoto}{Yu Christine Chen (S'10)} received the B.A.Sc. degree in Engineering Science (major in Electrical Engineering) from the University of Toronto in Canada in 2009 and the M.S. degree in Electrical Engineering from the University of Illinois at Urbana-Champaign in 2011. She is currently pursuing a Ph.D. degree in Electrical Engineering at the University of Illinois at Urbana-Champaign. Her research interests include power system dynamics and monitoring, and renewable resource integration.
%\end{IEEEbiographynophoto}
%
%\begin{IEEEbiographynophoto}{Sairaj V. Dhople (S'09--M'13)} received the B.S., M.S., and Ph.D. degrees in electrical engineering from the University of Illinois, Urbana-Champaign, in 2007, 2009, and 2012, respectively. He is currently an Assistant Professor in the Department of Electrical and Computer Engineering at the University of Minnesota, Minneapolis, MN, USA, where he is affiliated with the Power and Energy Systems research group. His research interests include modeling, analysis, and control of power electronics and power systems with a focus on renewable integration. Dr. Dhople received the NSF CAREER award in 2015. He currently serves as an Associate Editor for the IEEE Transactions on Energy Conversion.
%\end{IEEEbiographynophoto}
\end{document}